\documentclass[aps,prl,twocolumn,superscriptaddress,longbibliography]{revtex4-2}
\usepackage{amsmath}
\usepackage{graphicx}
\usepackage{hyperref}
\usepackage{color}
\usepackage{amsfonts}
\usepackage{amssymb}
\usepackage{braket}
\usepackage{times}
\usepackage{natbib}
\usepackage{lineno}
\usepackage{siunitx}

\begin{document}

\title{ \huge Broadband Spontaneous Parametric Downconversion in Reconfigurable Poled Linearly-Uncoupled Resonators}

\author{Alessia Stefano}
\email{alessia.stefano01@universitadipavia.it}
\affiliation{Dipartimento di Fisica, Università di Pavia, via Bassi 6, 27100 Pavia, Italy}

\author{Luca Zatti}
\affiliation{Dipartimento di Ingegneria Industriale e dell’Informazione, Università di Pavia, Via A. Ferrata 5, 27100 Pavia, Italy}

\author{Marco Liscidini}
\affiliation{Dipartimento di Fisica, Università di Pavia, via Bassi 6, 27100 Pavia, Italy}
\date{June 26, 2024}

\begin{abstract}
In this letter, we study spontaneous parametric down-conversion (SPDC) in a periodically poled structure composed of two linearly uncoupled resonators that are nonlinearly coupled via a Mach-Zehnder interferometer. The device does not require dispersion engineering to achieve efficient doubly-resonant SPDC and, unlike the case of a single resonator, one can reconfigure the system to generate photon pairs over a bandwidth of hundreds of nm. We consider the case of SPDC pumped at 775 nm in a periodically poled lithium-niobate (PPLN) device compatible with up-to-date technological platforms. We demonstrate pair generation rates of up to 250 MHz/mW pump power for a single resonance and integrated pair generation rates of up to 100 THz/mW pump power over 170 nm. When properly reconfigured, a single device can efficiently generate over a bandwidth of some 300 nm, covering the S, C, L, and U infrared bands. 

\end{abstract}

\maketitle

Photon pair generation holds significant promise for applications in quantum information processing and quantum communication \cite{flamini2018photonic,couteau2023applications}. Nonclassical light, which exhibits unique quantum characteristics such as entanglement and squeezing, can be generated through nonlinear processes like spontaneous four-wave mixing (SFWM) and spontaneous parametric down-conversion (SPDC). The efficiency of these nonlinear interactions can be significantly enhanced by spatial and temporal confinement within integrated optical resonators \cite{helt2012does}. In particular, micro-ring resonators offer a platform for achieving high-quality factors (Q) and small mode volumes, which are essential for enhancing the nonlinear interaction strength. Traditionally, SPDC is typically used in bulk systems \cite{kwiat1999ultrabright}, while SFWM is the primary choice for integrated devices \cite{caspani2017}.
Yet, recent advances in material science, such as the development of low-loss waveguides \cite{zhang2017monolithic} and hybrid integration techniques \cite{schuhmann2023hybrid}, have further improved the performance of SPDC-based integrated photon pair sources. In this respect, thin-film lithium niobate (TFLN) is a very appealing material platform, for it offers several advantages, including a relatively large second-order nonlinearity, large electro-optic coefficient, and low losses \cite{vazimali2022applications,zhu2021integrated}. Over the past decade, significant efforts have been directed towards optimizing the fabrication of TFLN waveguides \cite{desiatov2019ultra}. Both resonant \cite{lu2019periodically, bruch2019chip} and non-resonant \cite{rao2019actively} structures have been investigated for second-harmonic generation, demonstrating high-efficiency processes. Additionally, SPDC has been successfully implemented in TFLN waveguides \cite{zhao2020high,chen2019efficient,javid2021ultrabroadband}, achieving high generation rates with minimal losses.

\begin{figure}[t]
\includegraphics[width=\linewidth]{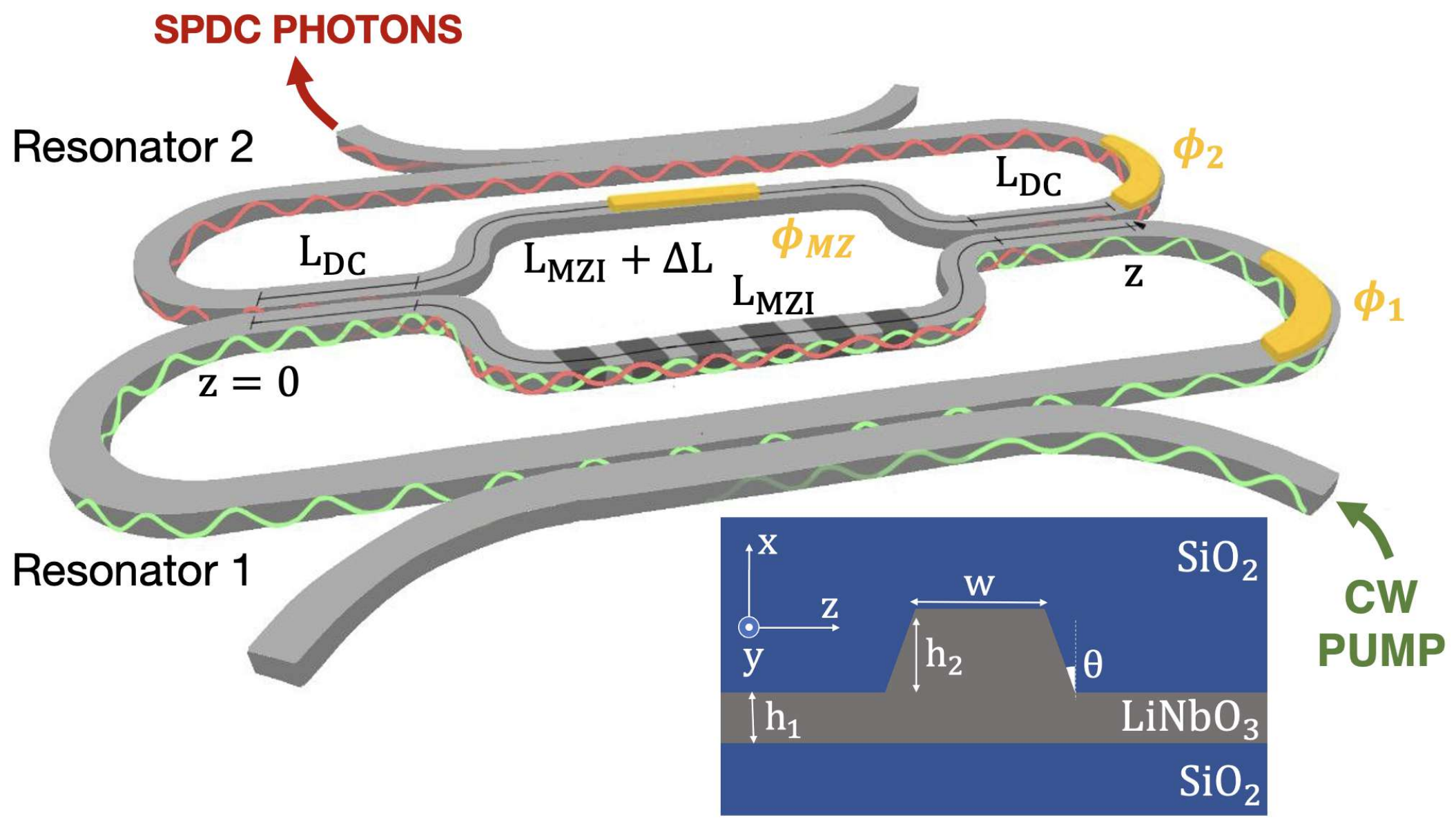}
\caption{Structure of the device. The inset shows the cross section of the waveguide, where $w=\SI{900}{nm}$, $h_2=\SI{400}{nm}$, $h_1=\SI{300}{nm}$, and $\theta=\ang{30}$ (not to scale).}
\label{structure}
\end{figure}

The design of resonant devices for SPDC is usually challenging because one must deal with optical fields over a broad spectral range, for the pump photons have an energy $\hbar\omega_P$ that is twice the average energy of the generated ones at the frequencies $\omega_s$ and $\omega_i$. The first challenge is usually compensating for chromatic dispersion to achieve phase matching (PM) over the desired frequency range. In this respect, various strategies have been developed, including dispersion engineering \cite{zatti2022spontaneous}, cyclic quasi-PM \cite{yang2007enhanced}, and periodical poling (PP) in ferroelectric materials \cite{zhao2020high}. 
The second difficult task is designing a device to obtain resonant enhancement of all the fields involved in the interaction. This requires resonances at $\omega_p$, $\omega_s$, and  $\omega_i$ while simultaneously satisfying the energy conservation condition \( \omega_p = \omega_s + \omega_i \). Although second-order nonlinear processes in ring resonators have been demonstrated \cite{lu2019periodically}, the operation bandwidth is usually limited, as well as the reconfigurability of the device after fabrication, which makes the performances particularly sensitive to fabrication errors.

A possible approach to overcome such limitations makes use of devices comprising linearly uncoupled ring resonators that share a common region of space in which the nonlinear interaction can take place \cite{menotti2019nonlinear}. Recently, we proposed a system in which a simple directional coupler (DC) provides linear uncoupling of the two resonators and, at the same time, allows for the nonlinear interaction to occur \cite{zatti2022spontaneous}. In this device the generation of uncorrelated photon pairs is possible over a bandwidth that is more than two orders of magnitude larger than that of an equivalent single resonator. However, it still requires significant dispersion engineering and inter-modal phase matching.

In this work, we make a significant step forward by considering a different configuration in which linearly uncoupled resonators are obtained using a Mach-Zehnder interferometer (MZI) and phase matching is achieved by periodical poling. The combination of these two approaches boosts the performance of our device well beyond those demonstrated so far in either periodically poled rings or in linearly uncoupled resonators. The resulting structure does not require particular dispersion engineering, is reconfigurable over a much larger bandwidth, and guarantees high efficiency since the nonlinear interaction involves fundamental modes for both pump and generated photons.

\begin{figure}[t]
\centering
\includegraphics[width=\linewidth]{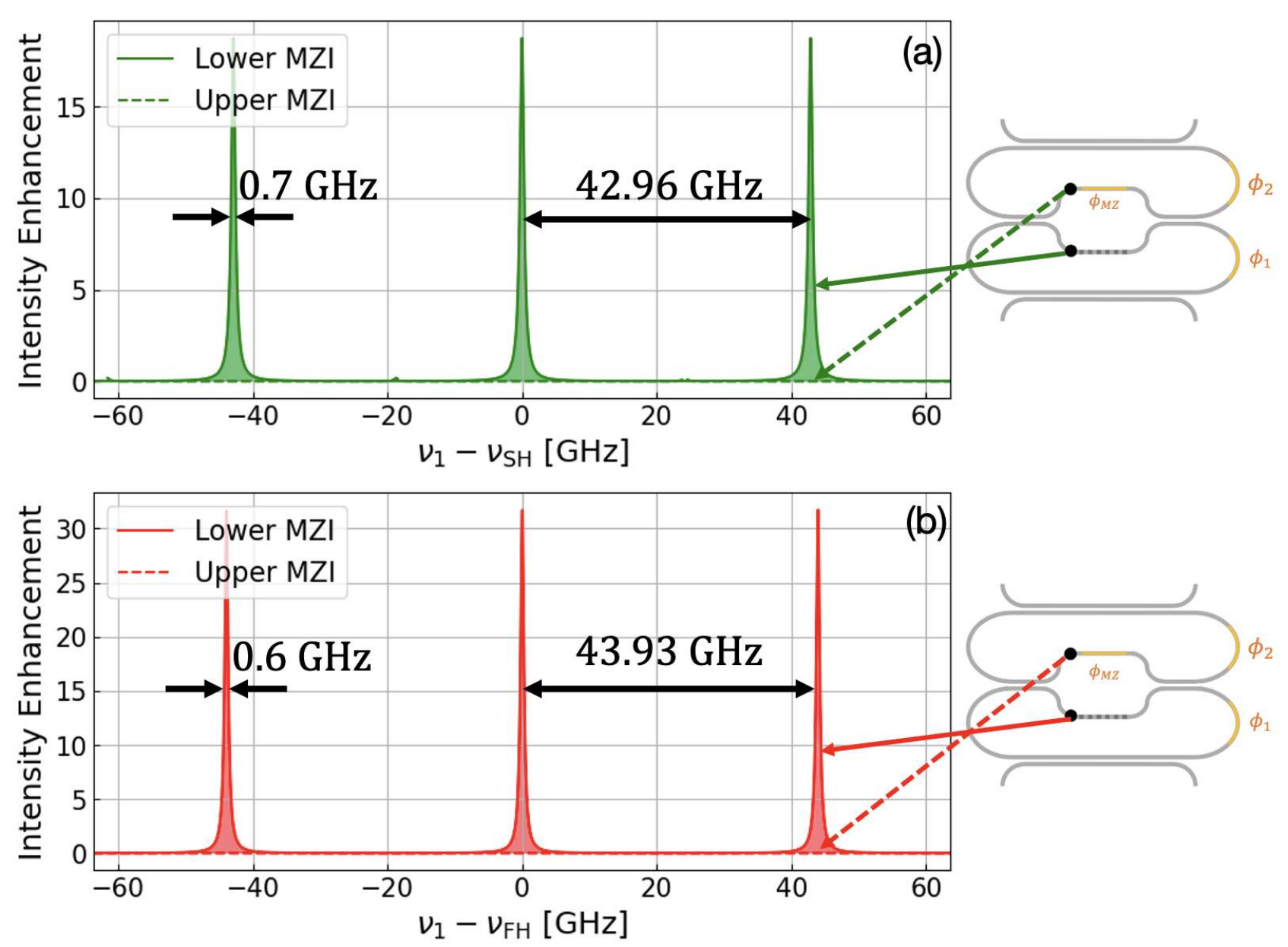}
\caption{Intensity enhancement in the upper (dashed line) and lower (solid line) arms of the MZI, as a function of the frequency $\nu_1$, centered around (a) $\nu_\mathrm{SH}$ and (b) $\nu_\mathrm{FH}$.}
\label{FE}
\end{figure}

A sketch of the device is shown in Fig \ref{structure}.  We consider a structure built from a \SI{700}{nm}-thick x-cut TFLN trapezoidal rib waveguide having a top width of $w=\SI{900}{nm}$, an etch depth of $h_2=\SI{400}{nm}$, and an etch angle of $\theta=\ang{30}$. We take SiO$_2$ as the upper and lower cladding material. We assume propagation losses of \SI{0.4}{dB/cm} at telecom wavelengths around the fundamental harmonic (FH) $\lambda_\mathrm{FH}= \SI{1550}{nm}$ and \SI{0.8}{dB/cm} around the second harmonic (SH) $\lambda_\mathrm{SH}= \SI{775}{nm}$ \cite{zhu2021integrated}. The group velocity dispersions (GVDs) are $\beta_{2,\mathrm{FH}}=\SI{3.6e-25}{s^2/m}$ at \SI{1550}{nm} and $\beta_{2,\mathrm{SH}}=\SI{7.0e-27}{s^2/m}$ at \SI{775}{nm}. The structure consists of two resonators shaped to form an MZI with two identical DCs. The resonators have a length of about \SI{3}{mm}, with a bending radius $R=\SI{60}{\micro m}$. The length $L_\mathrm{MZI}$ of the MZI arms is about \SI{906}{\micro m}, with the two arms slightly unbalanced of $\Delta L=\SI{0.4}{\micro m}$. The value of $\Delta L$ is chosen such that $k_\mathrm{FH}\Delta L = \pi$ to guarantee that the two resonators are linearly uncoupled over a range of some 300 nm centred at 1550 nm. In this spectral range, interference at the MZI output ensures that light entering the upper (lower) input port is entirely redirected in the upper (lower) output port, regardless of the splitting ratio in the DC. This feature enhances the robustness of our system against fabrication errors. \cite{sabattoli2022nonlinear,zatti2023generation}. The length of the DCs is \SI{179}{\micro m} to maximize cross transmission at 1550 nm and to minimize cross transmission at 775 nm. Finally, two DCs allow for the independent input/output coupling of each resonator, with a typical field distribution for the resonant modes of resonator 1 around 775 nm and resonator 2 in the infrared wavelength range sketched in Fig. \ref{structure}. We also consider the possibility of tuning the optical properties of the resonators and the MZI by means of three phase shifters $\phi_1 $, $\phi_2$, and $\phi_{MZI}$. Considering the chosen material, we imagine such phase shifters to be implemented with either thermo-optical or electro-optical modulators. 

In Fig.\ref{FE}, we plot the intensity enhancement of the electric field in the upper and lower arms of the MZI, for frequencies around $\nu_\mathrm{SH}=\SI{386}{THz}$ (\SI{775}{nm}) and $\nu_\mathrm{FH}=\SI{193}{THz}$ (\SI{1550}{nm}) for light injected from the lower and upper input port, respectively. Both plots feature two combs of resonances with quality factor $Q_\mathrm{SH}=10^5$ and $Q_\mathrm{FH}= 0.5 \times 10^5$ and with free spectral range (FSR) $\mathrm{FSR}_\mathrm{SH}= \SI{42.96}{GHz}$ and $\mathrm{FSR}_\mathrm{FH}=\SI{43.93}{GHz}$. The resonances linewidth are \SI{0.7}{GHz} for Resonator 2, and $\SI{0.6}{GHz}$ for Resonator 1. Propagation losses and the frequency dependence of the directional coupler, Mach-Zehnder interferometer and phase shifters are taken into account in our calculations. Here we designed the structure to have resonances at 775 nm and 1550 nm, for Resonator 1 and 2, respectively. As expected, in these frequency ranges, the fields are significant only in the lower arm of the MZI (solid lines in Fig.\ref{FE}).

An important features of our device is that the combs of resonances of the two resonators can be controlled independently by design. Indeed, one can set the FSRs of the two combs with a proper choice of the resonators' lengths. In addition, having two distinct coupling points allows one to reach different coupling condition for Resonator 1 and Resonator 2, ranging from undercoupling to overcoupling. A second important property is that, once the device has been fabricated, the relative position of the two combs and the effective coupling between Resonator 1 and Resonator 2 can be adjusted on-demand by changing the phases $\phi_1$, $\phi_2$, and $\phi_{\mathrm{MZI}}$.

The peculiar properties of our structure allows one to envision second-order nonlinear interaction between the two linearly uncoupled resonators. To this end, we consider the lower arm of the MZI to be periodically poled with a period $L_p=2\pi/(k_\mathrm{SH}-2k_\mathrm{FH})$ of 4.12 $\mu m$ to obtain Type-0 quasi-PM for SPDC involving the fundamental TE-modes. This allows one to exploit the strongest nonlinear coefficient in x-cut LN, $d_{33}=\SI{27}{pm/V}$. We assume to pump the system from Resonator 1 and collect the generated photon pairs at the output of Resonator 2 (see Fig. \ref{structure}).

In the hypothesis of a small generation probability, the state of the photon pair generated via SPDC in our system can be written as \cite{yang2008spontaneous}
\begin{equation}
    |II\rangle=\dfrac{1}{\sqrt{2}}\int d\omega_1 d\omega_2 \phi(\omega_1,\omega_2) a^\dagger_{\omega_1}a^{\dagger}_{\omega_2}|\mathrm{vac}\rangle,
\end{equation}
where $|\mathrm{vac}\rangle$ is the vacuum state, $a^\dagger_{\omega_i}$ is the creation operator for a photon at $\omega_i=2\pi \nu_i$ and
\begin{align}
\phi(\omega_1,\omega_2)=&i\dfrac{\alpha\bar{\chi_2}}{\beta}\sqrt{\dfrac{\hbar\omega_S\omega_I\omega_P}{8\pi\epsilon_0 n_In_Sn_Pc^3\mathcal{A}}}\nonumber\\
&\times \phi_P(\omega_1+\omega_2)J(\omega_1,\omega_2,\omega_1+\omega_2)
\end{align}
is the biphoton wavefunction. Here, $c$ is the speed of light, $\epsilon_0$ is the vacuum dielectric permittivity, $|\alpha|^2$ is the average number of photons per pump pulse, $|\beta|^2$ is the number of generated pairs per pump pulse (for $|\beta|^2\ll 1$), $\bar{\chi}_2$ is the typical value for the second-order nonlinear coefficient, and $\mathcal{A}$ is the nonlinear effective area \cite{yang2008spontaneous}. In addition, $n_P$, $n_S$ and $n_I$ are the effective refractive indices at the frequencies of the pump, signal and idler photons, respectively. Finally, $\phi_P(\omega)$ is the pump spectral profile, and $J(\omega_1,\omega_2,\omega_1+\omega_2)$ is the nonlinear field overlap integral \cite{introini2020spontaneous}, which can be written as
\begin{equation}
J(\omega_1,\omega_2,\omega_1+\omega_2)=f(\omega_1)f(\omega_2)f(\omega_1+\omega_2)I_{NL},
\end{equation}
where $f(\omega_i)$ are the field amplitudes in the lower arm of the MZI at $\omega_i$, and
\begin{align} \nonumber
    I_{NL}(\omega_1,\omega_2,\omega_1+\omega_2)= &\bar{\chi}_2\sum_{n=0}^N e^{in\Delta k L_p}\int_0^{L_p/2}e^{i\Delta k z} dz\\
    =\bar{\chi}_2 \dfrac{\sin(N \Delta k L_p/2)}{\sin(\Delta k L_p/2)} &\dfrac{L_p}{2}e^{i\Delta k L_p/4}\mathrm{sinc}(\Delta k L_p/4)
    \label{overlap}
\end{align}
takes into account the propagation of the fields along the $z$ direction (see Fig. \ref{structure}) and the effect of periodic poling. Here, $N=L_\mathrm{MZI}/L_p$ is the number of periods and $\Delta k =k(\omega_1+\omega_2)-k(\omega_1)-k(\omega_2)$. The overlap integral $J(\omega_1,\omega_2,\omega_1+\omega_2)$ plays a pivotal role in understanding the process of SPDC. It provides insights into the device's operating bandwidth and influences the generation rate.

\begin{figure}[t]
\centering
\includegraphics[width=\linewidth]{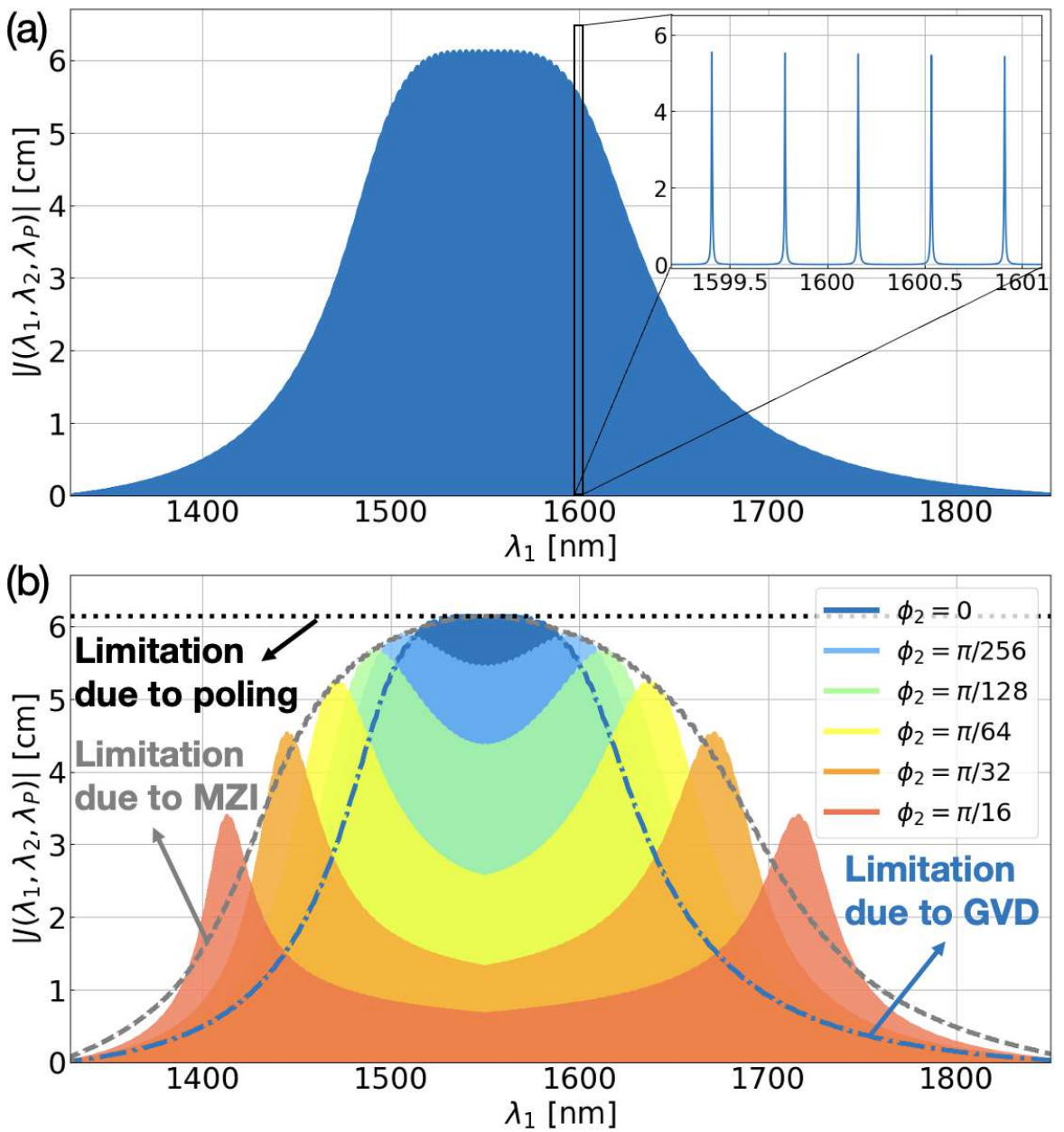}
\caption{(a) Overlap integral as a function of the wavelength $\lambda_1$ of one photon, in the case of CW pump ($\lambda_P=\SI{775}{nm}$ and $\lambda_2=\lambda_P\lambda_1/(\lambda_1-\lambda_P)$. Inset: zoom on a range near $\SI{1600}{nm}$. (b) Overlap integral with different additional phases $\phi_2$ between the resonators.}
\label{Overlap}
\end{figure}

We consider the case of a continuous wave (CW) pump centred at $\omega_P = \omega_\mathrm{SH}$, with $\omega_2=\omega_P-\omega_1$. In this scenario, the overlap integral is a function of the sole $\omega_1$ and can be written as $J(\omega_1,\omega_P-\omega_1,\omega_P)$. In Fig. \ref{Overlap}(a), we show the overlap integral as a function of the wavelength $\lambda_1=2\pi c/\omega_1$ of one of the generated photons for the structure under consideration. One can observe a dense frequency comb (see inset) whose envelope exhibits a plateau centred at 1550 nm, with a width of some 80 nm. In this spectral region the resonances involved in the SPDC process are symmetrically located with respect to $\omega_\mathrm{P}/2$. This means that resonant enhancement is achieved for signal, idler and pump fields, simultaneously. The envelope of the overlap integral gradually decreases to zero, resulting in a bandwidth of about \SI{170}{nm} over which the nonlinear interaction is resonantly enhanced. Such a bandwidth is limited by the GVD at \SI{1550}{nm}, which causes an increasing shift of the resonances from the optimal position as we move towards longer or shorter wavelengths. The effect of this shift on the overlap integral depends on the resonance linewidth $\lambda_\mathrm{FH}/Q_\mathrm{FH}$. The frequency bandwidth over which the shift is smaller than the resonance linewidth is $\Delta \omega_{\mathrm{GVD}} = 2 [\omega_1/(Q_F v_\mathrm{g,FH} \beta_\mathrm{2,FH})]^{1/2}$. 

By changing the phase $\phi_2$ of Resonator 2, it is possible to adjust the position of the comb of resonances and compensate the effect of the GVD to achieve resonant enhancement of SPDC in different spectral regions. This possibility is shown in Fig. \ref{Overlap}(b), in which we plot the overlap integrals for six values of $\phi_2$ from 0 to $\pi/16$. When $\phi_2$ is equal to 0, the overlap integral has the same shape shown in Fig.\ref{Overlap}(a). As the value of $\phi_2$ is increased, the maximum of the overlap integral splits in two new peaks which gradually move to larger and smaller wavelengths. Such peaks correspond to the spectral regions for which the structure is resonant for all the fields involved in the nonlinear interaction, thus maximizing their field enhancement simultaneously. 

There are two major effects that may limit the overlap integral and the efficacy of the reconfigurability: (i) the phase-matching bandwidth determined by the periodical poling; (ii) the frequency response of the MZI. Both these limitations are highlighted in Fig. \ref{Overlap}(b) with dashed lines. The phase-matching bandwidth is given by $\Delta \omega_{PM}=2[\pi/(L_p \beta_{2,FH}]^{1/2}$ and is much larger than the entire frequency range shown in Fig \ref{Overlap}, for which we verified that an expansion of the dispersion relation up to the second order around $\lambda_{\mathrm{FH}}$=\SI{1550}{nm} is accurate.
As it is evident from Fig.\ref{Overlap}(b), the frequency response of the MZI is the main limiting factor in our structure. In this specific example, we do not optimize the MZI to reach the largest bandwidth, and we just consider the simplest structure in which the two DCs are composed of two uniform parallel waveguides of length $L_\mathrm{DC}$. Alternative designs to increase the operation bandwidth for specific goals could be considered \cite{lin2021ultra}, but they are beyond the scope of this work. Interestingly, when $\phi_2=\pi/16$, we notice an additional enhancement of the overlap integral beyond the limitation of the MZI, around \SI{1400}{nm} and \SI{1700}{nm}. It turns out that in this configuration Resonator 1 and Resonator 2 are strongly coupled, and the resulting resonant splitting around \SI{1400}{nm} and \SI{1700}{nm}  helps to recover SPDC resonant enhancement.

\begin{figure}[t]
\centering
\includegraphics[width=\linewidth]{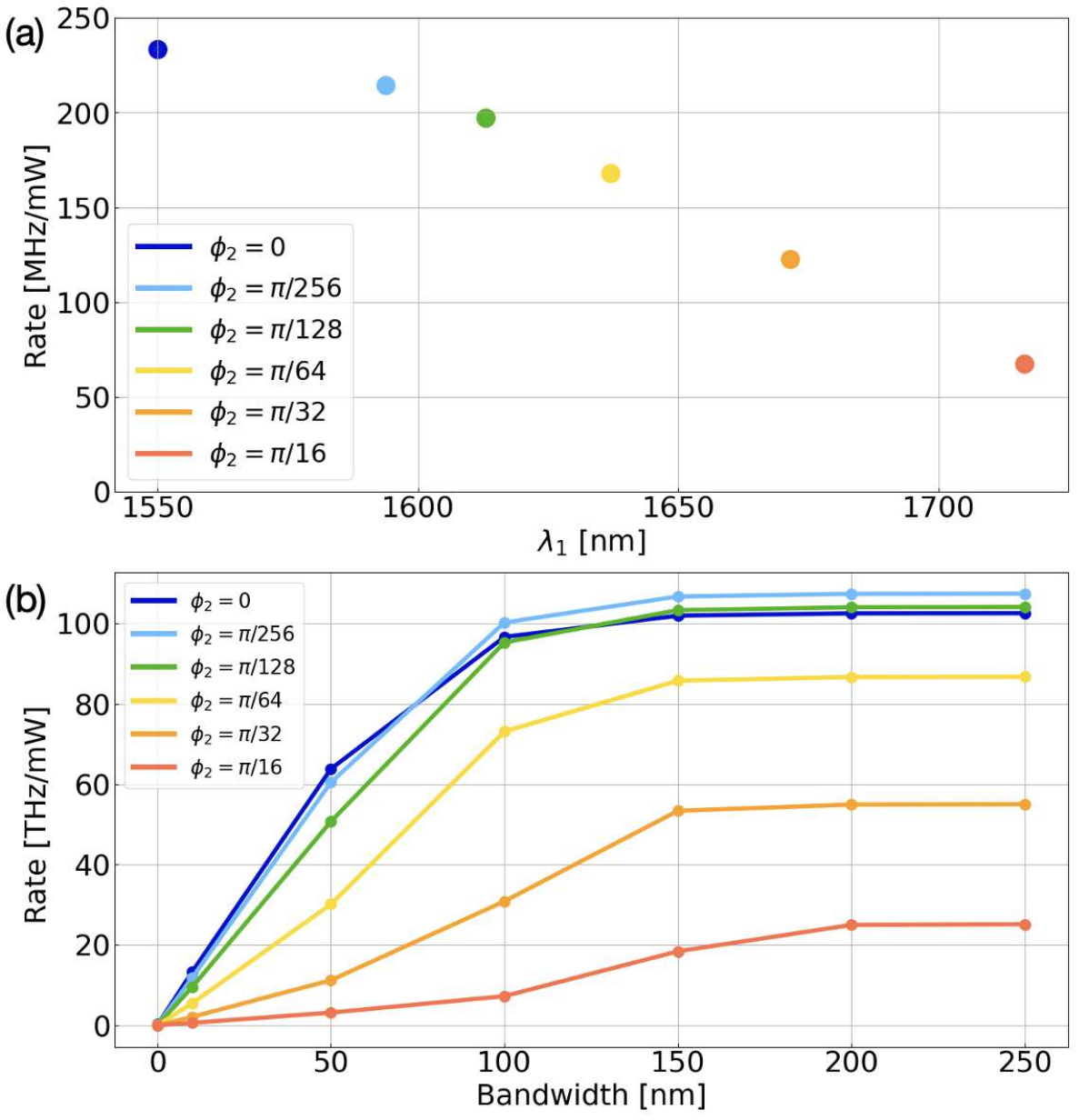}
\caption{(a) Pair generation rate for a single resonance, in the degenerate ($\Delta \lambda=0$) and non degenerate case ($\Delta \lambda\neq 0$). (b) Integrated pair generation rate over different bandwidth around $\lambda_{FH}$= \SI{1550}{nm}.}
\label{Rate} 
\end{figure}
We now examine the efficiency of our structure by calculating the pair generation rate. When considering a CW pump with power $P_P$, this can be written as \cite{onodera2016parametric}
\begin{align}
    R_\text{SPDC}^\text{CW} = &\frac{P_P \Bar{\chi}_{2}^2 \omega_S\omega_I}{8 \pi \varepsilon_0 c^3 \mathcal{A} \, n_{S}n_{I}n_{P}}
    \int |J(\omega,\omega_\text{P}-\omega, \omega_\text{P})|^2 d\omega .
\end{align}
First, we restrict our analysis by considering photon pairs generated in a single signal-idler pair of resonances. In Fig. \ref{Rate}(a), we plot the largest achievable rate/mW as a function of the wavelength of the center of signal resonance for the six values of $\phi_2$ considered in Fig. \ref{Overlap} (b). It should be noticed that, despite the nonlinear interaction occurs only in the lower MZI, it is possible to achieve generation rates exceeding hundreds of MHz/mW. This is determined by the resonant field enhancement provided by the two resonators independently, and by the simultaneous optimization of the coupling condition for pump, signal and idler fields \cite{banic2022two}. Finally, as expected, for larger wavelengths the generation rate decreases due to the limitation imposed by the frequency dependence of the MZI (see Fig. \ref{Overlap} (b)).

One of the main features of our device is the large generation bandwidth, which is particularly appealing in the context of spectral multiplexing \cite{joshi2018frequency}, or for the generation of high-dimensional states with frequency-bin encoding \cite{lu2023frequency,lu2018quantum}. In Fig \ref{Rate}(b) we show the integrated generation rate as a function of the bandwidth center at 1550 nm for the six values of $\phi_2$. Pair generation rates from 20 THz/mW to 100 THz/mW are achievable for the configurations under exam. Such values are limited by the MZI response and GVD, but further optimization is possible depending on the final application.

In conclusion, we proposed and investigated SPDC in linearly uncoupled resonators comprising of a PPLN MZI. We demonstrated that our device effectively addresses two primary challenges in resonant SPDC: achieving large phase-matching bandwidth and resonant enhancement for pump, signal, and idler fields. This is obtained by considering two different resonators, one for the pump and one for the generated photons, which share a common region of space in which SPDC can take place. Thus, one can control the relative position of the resonances at the pump and generation wavelengths without heavily relying on dispersion engineering. We show that by considering realistic structure parameters that are fully compatible with state of the art fabrication, a single device can operate over the S, C, L, and U infrared bands, with generation rates exceeding 50 MHz/mW, for a single pair of signal/idler resonances, and up to 100 THz/mW, when integrating over the entire generation bandwidth. This device holds promise for various applications, including entangled photon pair generation, heralded single-photon sources, and frequency-bin encoding. 

\bibliography{sample}

\end{document}